# Entanglement Transfer and C-Not Gate Implemented in A Fiber Connected 3-Atom System

Yao DENG, Yan-Qing GUO*, Pei PEI, Dian-Fu WANG, and Dong MI
Department of Physics, Dalian Maritime University
Dalian, P. R. China
*Corresponding author, yqguo@dlmu.edu.cn

*Abstract*—A robust scheme for unknown entangled state transfer and C-Not gate implemented between two spatially separated atoms is proposed. It is shown that, in the effective three-atom Ising model, two-atom unknown entangled state can be transferred from one pair of atoms to another deterministically by only repeating a simple operation of turning on/off local laser field applied on atom for two times at controlling time $\pi/2\Omega_0$. The whole time cost is less than $3\pi/2\Omega_0$. The successful probability and the fidelity are almost 100% for small atomic spontaneous emission rate. Deterministic two-atom C-Not gate can also be implemented in this model by just turning on/off the local laser field applied on the single target atom and leaving the controlling atom in its cavity alone for a while. The whole time cost is less than $\pi/2\Omega_0 + 3\pi\Delta/2g^2$. It is demonstrated that the scheme is insensitive to cavity leakage and fiber loss.

*Keywords- Entangled state Transfer, C-Not Gate, Ising Model, Fidelity*

## I. Introduction

Very recently, many studies focus on implementing unknown quantum state transfer [1-6], especially unknown entangled state, and controlled Not (C-Not) gate between distant qubits [6,7], since they play key roles for distant quantum communication and distributed quantum computation. Many schemes pay attention to the systems including atom-atom interaction that is created in fiber-connected cavity QED systems[2, 6, 8-10]. Most of the proposed schemes use multi-step geometric control including photon detection technic. This leads to the schemes working in a probabilistic way. To practically improve the success probability and the implementation fidelity, one must establish precisely controlled interaction between qubits and try to weaken the affect of coincidence detection inefficiency. It has been deeply understood that spin-spin coupling between two quantum dots is one of the very expected interaction that can be easily and accurately controlled by applying local electromagnetic fields without likelihood detection [1, 4, 5, 11-15]. While, the main obstacle in such schemes is how to precisely control individual spin because any kinds of local field controlling on a single spin may take affect on the very nearby neighboring ones. However, there are schemes put forward to construct effective spin-spin interaction between distant atoms by means of fiber-connection. In the scheme proposed by Mancini and Bose [8], an Ising-type interaction between two two-level atoms trapped in fiber-connected cavities is engineered. This model is extended to three-atom cases [16]. In another scheme suggested by Zhong et al [14], an effective Heisenberg spin chain is realized in a series of cavities connected by optical fibers. These studies open a new site for implementing scalable quantum communication and computation. In the present paper, we propose an alternative scheme based on a system consists of three distant intra-cavity atoms, which is proved to be an effective tripartite Ising model. We discuss the programming operation for implementing unknown two-atom entangled state transfer in two steps, controlled U (C-U) gate in one step and controlled Not gate in two steps. We demonstrate that the scheme works in high fidelity even the atomic spontaneous emission is included.

## II. Model Description and Secular Hamiltonian

The model setup of the global quantum system located in vacuum is shown in Fig. 1. Three single-sided Gaussian cavities are sequentially linked by optical fibers in ring connection. A two-level atom (with an excited state $|e\rangle$ and a ground state ) is trapped in each cavity and interacts with the single mode cavity field, which is driven by an external field, in dispersive way with large detuning $\Delta \gg g$, where $g$ is the coupling strength between atom and cavity field.

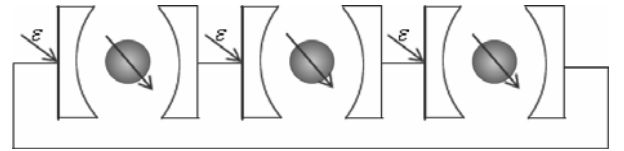

Figure 1. Schematic setup of the supposed system. Three two-level atoms are trapped in separate optical cavities linked via optical fibers. Each of the cavities is driven by an external field. Every atom is coupled to a local laser field.

Each intra-cavity atom interacts with a local weak laser field. This leads to a modified Jenneys-Cummings model [17] under rotating wave approximation (RWA) with effective Hamiltonian in interaction picture as

$$H_{JC} = \chi|e\rangle\langle e| + \chi\sigma^z a^+ a H_{drv} + \varepsilon(a + a^+) + \Omega(\sigma + \sigma^+), \quad (1)$$

where $\chi = g^2/\Delta$, $\Delta$ is the detuning between atomic internal transition frequency and cavity field frequency, $a(a^+)$ is annihilation (creation) operator of the cavity field, $\sigma^z$ is the spin-Z operator of the atom, $\Omega$ is the Rabi frequency, the first term of the equation is the stark shift.

Assuming the cavity is connected with an optical fiber, and considering the leakage of the cavity, one can write the kinetic equation, namely the Heisenberg equation, of cavity operator under inputting-outputting condition introduced in [18]. Then one can obtain the value of operator of the cavity field fluctuating around its steady state by taking adiabatic approximation. Finally, we obtain the effective Hamiltonian of the global system in the interaction picture [16], by just substituting field operators into Eq. (1) and keeping terms including $\chi$ to the second order, as

$$H_{eff} = J(\sigma_1^z \sigma_2^z + \sigma_2^z \sigma_3^z + \sigma_1^z \sigma_3^z) + \sum_{i=1}^{3} \Omega_i (\sigma_i^+ + \sigma_i^-), \quad (2)$$

where $J$ is the atomic spin-spin coupling strength of effective Ising model. The Rabi frequency satisfies $\Omega_i \ll J$, and $J = 2\kappa\chi^2 \text{Im}\{|\alpha|^2 (Me^{i\phi} + \kappa e^{i2\phi})/M^3 - W^3\}$, where $\kappa$ is the rate of the cavity field leaking to the fiber, $M = i\Delta + \kappa$, $W = \kappa^3 e^{i3\phi}$, with $\phi$ the phase delay of the photon transmitting from one cavity to another along a single fiber, $\alpha$ is the steady value of cavity field operator and satisfies $\alpha = \varepsilon(M^2 + M\kappa e^{i\phi} + k^2 e^{i2\phi})/M^3 - W^3$.

We can take further approximation for Eq. (2) under RWA by using a transformation

$$\tilde{H} = U^{-1}[\sum_{i=1}^{3} \Omega_i (\sigma_i^+ + \sigma_i^-)]U \quad (3)$$

where $U = e^{-iH_{zz}t}$, $H_{zz} = J(\sigma_1^z \sigma_2^z + \sigma_2^z \sigma_3^z + \sigma_1^z \sigma_3^z)$. One can calculate and simplify the transformation by substituting the equality $U = \sum_{i \neq j} \cos Jt - i\sigma_i^z \sigma_j^z \sin Jt$ into Eq.(3), and obtain the secular part of the effective Hamiltonian by neglecting the fast oscillating terms including $\cos Jt$ or $\sin Jt$ under the condition $\Omega_i \ll J$. The secular Hamiltonian [15, 19] can be written as

$$H_{secular} = \sum_{ijk} \Omega_i \sigma_i^x (1 - \sigma_j^z \sigma_k^z) \quad (4)$$

where the subscript numbers $i, j, k$ are permutations of 1, 2, 3 in turn.

## III. DETERMINISTIC ENTANGLEMENT TRANSFER UNDER GEOMETRIC CONTROL

The task for entanglement transfer from subsystem 1, 2 to subsystem 2, 3 in a tripartite two-level quantum system is to implement the operation [6]

$$|\Psi\rangle_{12} \otimes |g\rangle_3 \rightarrow |g\rangle_1 \otimes |\Psi\rangle_{23} \quad (5)$$

where $|\Psi\rangle_{12}$ and $|\Psi\rangle_{23}$ are entangled states for subsystem 1, 2 and subsystem 2, 3 respectively. In a quantum dynamical process under geometric control, the former in Eq. (5) turns out to be the initial state of the system and the latter is the target state.

In this model, we assume the initial system state is $(\alpha|e\rangle_1|g\rangle_2 + \beta|g\rangle_1|e\rangle_2)|g\rangle_3$ or $(\alpha|e\rangle_1|e\rangle_2 + \beta|g\rangle_1|g\rangle_2)|g\rangle_3$, the target state is $|g\rangle_1(\alpha|e\rangle_3|g\rangle_2 + \beta|g\rangle_3|e\rangle_2)$ or $|g\rangle_1(\alpha|e\rangle_3|e\rangle_2 + \beta|g\rangle_3|g\rangle_2)$. $\alpha$ and $\beta$ are unknown normalized factors.

We now analyze the evaluated system state under the condition $\Omega_1 = \Omega_0$, $\Omega_2 = \Omega_3 = 0$. It can be easily understood that the state of atoms 2 and 3 keeps unchanged. The system state $|\psi(t)\rangle$ is dominated by the secular Hamiltonian of the system and can be obtained by solving Schrödinger equation

$$\frac{d}{dt}|\psi(t)\rangle = -iH_{secular}|\psi(t)\rangle \quad (6)$$

For initial state $|e\rangle_1|g\rangle_2|g\rangle_3$, the system state keeps unchanged because the spins of atom 2 and atom 3 are same, while for initial state $|g\rangle_1|e\rangle_2|g\rangle_3$, the system state is restricted within the Hilbert subspace spanned by the vectors $|\phi_1\rangle = |g\rangle_1|e\rangle_2|g\rangle_3$, $|\phi_2\rangle = |e\rangle_1|e\rangle_2|g\rangle_3$.

The secular Hamiltonian can be rewritten as

$$H_{secular} = \begin{bmatrix} 0 & \Omega_0 \\ \Omega_0 & 0 \end{bmatrix} \quad (7)$$

The evaluated system state can be obtained as

$$|\psi(t)\rangle = \cos\Omega_0 t |g\rangle_1|e\rangle_2|g\rangle_3 - i\sin\Omega_0 t |e\rangle_1|e\rangle_2|g\rangle_3 \quad (8)$$

Thus, for initial state $|\psi(0)\rangle = (\alpha|e\rangle_1|g\rangle_2 + \beta|g\rangle_1|e\rangle_2)|g\rangle_3$, the evaluated system state is

$$|\psi(t)\rangle = \alpha|e\rangle_1|g\rangle_2|g\rangle_3 + \beta(\cos\Omega_0 t |g\rangle_1|e\rangle_2|g\rangle_3 \\ -i\sin\Omega_0 t |e\rangle_1|e\rangle_2|g\rangle_3) \quad (9)$$

At time $t = \pi/2\Omega_0$, one can turn off the local laser field that applied on atom 1 to deterministically obtain a system state

$$|\psi(\pi/2)\rangle = \alpha|e\rangle_1|g\rangle_2|g\rangle_3 - i\beta|e\rangle_1|e\rangle_2|g\rangle_3 \quad (10)$$

Based on the above results, we program the whole route of the operating process for transferring entanglement from atoms 1 and 2 to atoms 2 and 3 as in TABLE I. In the table, the '$\pi/2$ pulse' is used to equivalently denote interacting time $t = \pi/2\Omega_0$.

It can be easily seen that the task assumed in Eq. (5) has been accomplished with time cost $t_{p_1} = 3\pi/2\Omega_0$. If $|g\rangle_1$ is not necessarily required as the outputting state of atom 1, the time cost of the operating process can be saved to $\pi/\Omega_0$. In Table 1, the operation process for transferring unknown entangled state does not require any coincidence measurement or any probabilistic outputting results. Thus, the successful probability and theoretical fidelity for this programming are both 100%. Similar operation process can be programmed for transferring unknown entangled state between atoms 2 and 3, and atoms 1 and 3, with the same time cost in this model.

TABLE I. OPERATING PROCESS FOR TRANSFERRING UNKNOWN ENTANGLED STATE BETWEEN DISTANT ATOMS 1 AND 2

| Operating sequence | Evaluated system state |
|---|---|
| $\Omega_1 = \Omega_0, \Omega_2 = \Omega_3 = 0$ | $(\alpha|e\rangle_1|g\rangle_2 + \beta|g\rangle_1|e\rangle_2)|g\rangle_3$ (inputting initial state) |
| '$\pi/2$ pulse' on atom 1 | $\alpha|e\rangle_1|g\rangle_2|g\rangle_3 - i\beta|e\rangle_1|e\rangle_2|g\rangle_3$ |
| $\Omega_3 = \Omega_0, \Omega_1 = \Omega_2 = 0$ | $\alpha|e\rangle_1|g\rangle_2|g\rangle_3 - i\beta|e\rangle_1|e\rangle_2|g\rangle_3$ |
| '$\pi/2$ pulse' on atom 3 | $|e\rangle_1(\alpha|g\rangle_2|e\rangle_3 + \beta|e\rangle_2|g\rangle_3)$ |
| $\Omega_1 = \Omega_0, \Omega_2 = \Omega_3 = 0$ | $|e\rangle_1(\alpha|g\rangle_2|e\rangle_3 + \beta|e\rangle_2|g\rangle_3)$ |
| '$\pi/2$ pulse' on atom 1 | $|g\rangle_1(\alpha|g\rangle_2|e\rangle_3 + \beta|e\rangle_2|g\rangle_3)$ (outputting target state) |

In deducing the effective Ising-type Hamiltonian, the cavity field leakage is assumed to be large enough. So, this scheme is undoubtedly insensitive to cavity leakage. While, the atomic spontaneous emission which is inevitable in any quantum process can still take affect on the efficiency of the scheme and result in a dissipative dynamical evaluation. To quantitatively illustrate the influence of atomic spontaneous emission rate on average fidelity, one can rewrite the global Hamiltonian by adding a non-Hermitian conditional term [9] to the Hamiltonian in Eq. (7) as

$$H_s = -i\Gamma \sum_i |e\rangle_i \langle e| + H_{\text{secular}} \quad (11)$$

Motivated by the definition of the Hilbert-Schmidt distant $\|\rho(t) - \rho_d\|$ between two density operators [20], we define the average fidelity [6] as

$$F = \frac{1}{2\pi}\int_0^{2\pi}[1 - \frac{Tr(\rho(t)-\rho_d)^2}{2}]d\theta \quad (12)$$

To calculate the average fidelity, the coefficients $\alpha$ and $\beta$ in the initial state are replaced by $\cos\theta$ and $\sin\theta$. The average fidelity at controlling time $t_1 = t_2 = \pi/2\Omega_0$ is shown in Fig. 2.

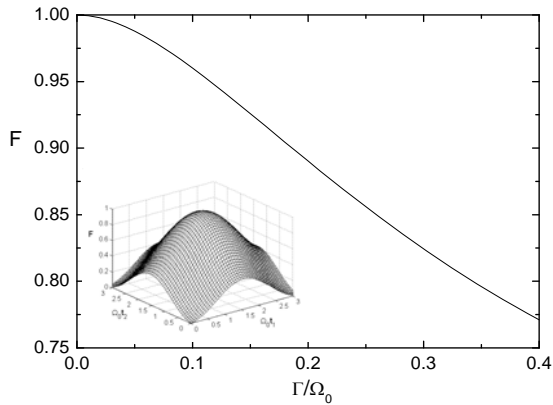

Figure 2. Schematic setup of the supposed system. Three two-level atoms are trapped in separate optical cavities linked via optical fibers. Each of the cavities is driven by an external field. Every atom is coupled to a local laser field.

Obviously, the average fidelity decreases slowly for relative weak atomic spontaneous emission rate. For $\Gamma = 0.05\Omega_0$, the average fidelity is $F = 0.99$, while for $\Gamma = 0.1\Omega_0$ the average fidelity is $F = 0.96$. One can conclude that the scheme works well if the atomic spontaneous emission rate is relative not large.

## IV. DETERMINISTIC C-NOT GATE BETWEEN TWO ATOMS

Moreover, this model can be applied to construct deterministic C-Not gate between two distant qubits a and b. For this purpose, the following implementation must be accomplished [6]

$$\begin{aligned}|e\rangle_a|e\rangle_b &\to |e\rangle_a|g\rangle_b \\ |e\rangle_a|g\rangle_b &\to |e\rangle_a|e\rangle_b \\ |g\rangle_a|e\rangle_b &\to |g\rangle_a|e\rangle_b \\ |g\rangle_a|g\rangle_b &\to |g\rangle_a|g\rangle_b\end{aligned} \quad (13)$$

To this end, we assume atom 1 is controlling qubit, atom 2 is controlled qubit, atom 3 turns out to be an auxiliary qubit and is initially prepared in ground state, and laser fields applied on atom 1 and atom 3 are zeros, which lead to the Rabi frequencies $\Omega_2 = \Omega_0$, $\Omega_1 = \Omega_3 = 0$. For initial system state $|\psi(0)\rangle = |e\rangle_1|e\rangle_2|g\rangle_3$, one can turn off the laser field acting on atom 2 at specific time $t = \pi/2\Omega_0$ and obtain the outputting system state $|\psi(t)\rangle = -i|e\rangle_1|g\rangle_2|g\rangle_3$. For initial system state $|\psi(0)\rangle = |e\rangle_1|g\rangle_2|g\rangle_3$, one can repeat the same operation and obtain the outputting system state $|\psi(t)\rangle = -i|e\rangle_1|e\rangle_2|g\rangle_3$. While for initial state $|\psi(0)\rangle = |g\rangle_1|e\rangle_2|g\rangle_3$ or $|\psi(0)\rangle = |g\rangle_1|g\rangle_2|g\rangle_3$, the outputting state is same as the inputting state because the neighboring atoms of atom 2 have same spins. However, there is a localized phase factor $-i$ that can not be directly neglected in the outputting system states for initial atom state $|e\rangle_1$. The quantum gate turns out to be a C-U gate, which is a combination of two-qubit C-Not gate and a single-qubit phase shift gate [7]. On one hand, it is well known that this kind of C-U gate is essential for constructing quantum computer and quantum network since any quantum logical computing process can be theoretically simulated to an arbitrary degree of accuracy through a quantum circuit combined only with C-Not gates and single qubit rotating operations. On the other hand, it has been demonstrated that the phase factor $-i$ can be eliminated by introducing another quantum electromagnetic field interacting with atom [7].

In this model, we recall that there exists a stark-shift term in the effective J-C model in Eq. (1). The phase factor $-i$ can be eliminated by leaving atom 1 alone in the cavity for a while after the above operations. The whole route of the operating process for deterministic C-Not Gate between atoms 1 and 2 can be programmed as in Tab. II.

The time cost for implementing C-Not gate is $t_{p_2} = \pi/2\Omega_0 + 3\pi\Delta/2g^2$. While for the implementation of C-$\pi/2$ gate (the single-qubit phase shift gate for rotating angle $\theta = \pi/2$), the time cost can saved to $\pi/2\Omega_0$.

TABLE II. OPERATING PROCESS FOR IMPLEMENTING C-NOT GATE BETWEEN ATOMS 1 AND 2

| Operating sequence | Evaluated system state |
|---|---|
| Atoms enter cavities. $\Omega_2 = \Omega_0$, $\Omega_1 = \Omega_3 = 0$ | $|e\rangle_1|e\rangle_2|g\rangle_3$ or $|e\rangle_1|g\rangle_2|g\rangle_3$ or $|g\rangle_1|e\rangle_2|g\rangle_3$ or $|g\rangle_1|g\rangle_2|g\rangle_3$ (inputting initial state) |
| '$\pi/2$ pulse' on atom 2 | $-i|e\rangle_1|g\rangle_2|g\rangle_3$ or $-i|e\rangle_1|e\rangle_2|g\rangle_3$ or $|g\rangle_1|g\rangle_2|g\rangle_3$ or $|g\rangle_1|g\rangle_2|g\rangle_3$ |
| Atoms 2 and 3 leave cavities. $\Omega_1 = \Omega_2 = \Omega_3 = \varepsilon = 0$ | $-i|e\rangle_1|g\rangle_2|g\rangle_3$ or $-i|e\rangle_1|e\rangle_2|g\rangle_3$ or $|g\rangle_1|e\rangle_2|g\rangle_3$ or $|g\rangle_1|g\rangle_2|g\rangle_3$ |
| '$3\pi/2$ pulse' on atom 1 | $|e\rangle_1|g\rangle_2|g\rangle_3$ or $|e\rangle_1|e\rangle_2|g\rangle_3$ or $|g\rangle_1|e\rangle_2|g\rangle_3$ or $|g\rangle_1|g\rangle_2|g\rangle_3$ |
| Atom 1 leaves cavity. | $|e\rangle_1|g\rangle_2|g\rangle_3$ or $|e\rangle_1|e\rangle_2|g\rangle_3$ or $|g\rangle_1|e\rangle_2|g\rangle_3$ or $|g\rangle_1|g\rangle_2|g\rangle_3$ (outputting target state) |

The '$3\pi/2$ pulse' in Tab. II is used to equivalently denote interacting time $t = 3\pi\Delta/2g^2$.

The operation in Tab. II can be concluded as the transformation $|\psi\rangle_{\text{output}} = U_{\text{C-Not}}|\psi\rangle_{\text{input}}$ with transforming matrix $U_{\text{C-Not}} = \begin{pmatrix} 1 & 0 & 0 & 0 \\ 0 & 1 & 0 & 0 \\ 0 & 0 & 0 & 1 \\ 0 & 0 & 1 & 0 \end{pmatrix}$, on basis of $|g\rangle_1|g\rangle_2|g\rangle_3$, $|g\rangle_1|e\rangle_2|g\rangle_3$, $|e\rangle_1|g\rangle_2|g\rangle_3$, and $|e\rangle_1|e\rangle_2|g\rangle_3$. This is exactly the C-Not gate between two qubits with theoretical successful probability 100% and theoretical fidelity 100%. Similar operations can be programmed to implement C-Not gate between atoms 2 and 3, and atoms 1 and 3 in this model.

## V. SUMMARY

In summary, we have proposed a deterministic unknown entangled state transfer scheme for a system contains three distant atoms. A geometric control is used to obtain the target entangled state deterministically. The only required operation is turning on/off the local laser fields applied on atoms that assumed to transfer the entangled state consequently at controlling time $\pi/2\Omega_0$. The average fidelity of transferring the unknown entangled state is 100%. We discuss the affect of atomic spontaneous emission on the fidelity. It is shown that the atomic spontaneous emission decreases the quantity of fidelity, while the scheme still works well and turns out to be robust for relative not large atomic spontaneous emission rate, especially for $\Gamma \leq 0.1\Omega_0$. It has been demonstrated that the dissipation of the photon leakage along optical fibers can be included in the spin-spin coupling strength by replacing $e^{i\phi}$ with $e^{i\phi - \gamma L}$ [8], where $\gamma$ is the fiber decay per meter and $L$ is the length of the optical fiber. The decay per meter is about $\gamma = 0.08$ for typical fibers [21]. The spin-spin coupling strength is now about $0.9J$, which ensures the conditions for adiabatic approximation and RWA, and keeps the derivation of secular part of effective Hamiltonian valid. The scheme is then insensitive to the variety of fiber leakage rate in strong leakage region. We also discussed the implementation of C-Not gate using this model. The only required operations for C-Not gate are turning on the local laser field applied on the controlled atom and turning off the laser field and the driven fields at controlling time $\pi/2\Omega_0$, and leaving atom 1 alone in the cavity for time $t = 3\pi\Delta/2g^2$. The stark shift, which is directly produced by the large detuning between the cavity and the intra-cavity atom without introducing any other auxiliary atom-cavity interaction, is used to rotate the state of atom 1 to eliminate the phase factor $-i$. The advantage of this scheme is that one can individually control the spin of the target atom by applying local laser field on the atom without disturbing any other atoms. Furthermore, one can design Lyapunov control to eliminate the time-delay affect inevitably caused by the mismatch of practical and theoretical controlling times [10, 20, 22].

## ACKNOWLEDGMENTS

We thank Professor Dian Min Tong for helpful discussions and his encouragement. This research was financially supported by NSF of China under grant NO. 11305021 and the Fundamental Research Funds for the Central Universities.